# LABORATORY INVESTIGATIONS OF THE INTERACTION BETWEEN BENZENE AND BARE SILICATE GRAIN SURFACES

J. D. T$_{\text{HROWER}}$[1]*, M. P. C$_{\text{OLLINGS}}$[1], F. J. M. R$_{\text{UTTEN}}$[2] AND M. R. S. M$^{\text{C}}$C$_{\text{OUSTRA}}$[1]

[1]School of Engineering and Physical Sciences, Heriot-Watt University, Edinburgh, EH14 4AS, UK.

[2]School of Pharmacy and iEPSAM, Keele University, Keele, ST5 5BG, UK

# ABSTRACT

Experimental results on the thermal desorption of benzene ($C_6H_6$) from amorphous silica ($SiO_2$) are presented. The amorphous $SiO_2$ substrate was imaged using atomic force microscopy (AFM), revealing a surface morphology reminiscent of that of interplanetary dust particles (IDPs). Temperature programmed desorption (TPD) experiments were conducted for a wide range of $C_6H_6$ exposures, yielding information on both $C_6H_6$–$SiO_2$ interactions and the $C_6H_6$-$C_6H_6$ interactions present in the bulk $C_6H_6$ ice. The low coverage experiments reveal complicated desorption behaviour that results both from porosity and roughness in the $SiO_2$ substrate, and repulsive interactions between $C_6H_6$ molecules. Kinetic parameters were obtained through a combination of direct analysis of the TPD traces and kinetic modelling, demonstrating the coverage dependence of both desorption energy and pre-exponential factor. Experiments were also performed whereby the pores were blocked by pre-exposure of the $SiO_2$ to water vapour. $C_6H_6$ was observed to be adsorbed preferentially on the $SiO_2$ film not covered by $H_2O$ at the temperature at which these experiments were performed. This observation means that intermolecular repulsion likely becomes important at smaller $C_6H_6$ exposures on grains with a $H_2O$ mantle. Kinetic modelling of $C_6H_6$ multilayer desorption yields kinetic parameters in good agreement with previous studies, with the $SiO_2$ having little impact on the desorption beyond the first few layers.

# 1. INTRODUCTION

Carbon is thought to exist in many forms throughout the interstellar medium (ISM). In particular, polycyclic aromatic hydrocarbons (PAHs) are generally considered to be one the most important classes of carbon bearing molecule (Ehrenfreund & Sephton 2006). PAHs have been proposed as carriers of both the diffuse interstellar bands (DIBs) (Duley 2006, and references therein) and the unidentified infrared bands (UIRs) (Allamandola, Hudgins & Sandford 1999). Absorption features corresponding to the UIRs have been observed towards several protostellar objects and towards the galactic centre (Bregman & Temi 2001, and references therein). PAH emission features have been observed in a wide range of environments, (see e.g. Kaneda, Onaka & Sakon 2005 ; Lagadec et al. 2005 ; Kassis et al. 2006 ; Flagey et al. 2006 ; Smith et al. 2007), demonstrating their ubiquitous nature. As with many gas-phase molecules, it is likely that PAHs are also present mixed in the water ($H_2O$) ice dominated mantles surrounding grains in the dense interstellar medium under suitably cold conditions. Laboratory spectra of PAHs in appropriate ice matrices are essential in aiding the interpretation of observational spectra. To this end there have been several experimental studies of the infrared absorption spectra of PAH/$H_2O$ ice mixtures (Sandford, Bernstein & Allamandola 2004 ; Bernstein, Sandford, Allamandola 2005 ; Bernstein et al. 2007).

We are currently undertaking work to investigate the irradiation of PAHs with both photons and electrons. As a first step to understanding the relevant processes, we are focussing on benzene ($C_6H_6$), primarily for reasons of experimental convenience, before considering simple PAH species and their heterocyclic analogues, including pyridine ($C_5H_5N$). As UV absorption characteristics of small PAHs and $C_6H_6$ are similar, any knowledge obtained from studies of this simpler system will contribute positively to our understanding of processes involving PAH larger species. Furthermore to aid in interpreting the results of irradiation experiments, it is necessary to understand the nature of the pre-irradiated ice. For example, in our recently reported experimental measurements of the photodesorption of $C_6H_6$ from $H_2O$ ice (Thrower et al. 2008), the desorption of both $C_6H_6$ and $H_2O$ was found to depend strongly on the morphology of the ices. It is clear that a more detailed understanding of the interactions between both $C_6H_6$ and $H_2O$ and the underlying substrate is required. As far as we are aware, there have been no experimental studies of the interaction between $C_6H_6$ and suitable grain mimics. We have chosen a grain mimic based on a thin film of amorphous silica in order to represent the silicate grain population. Compared to our previous study of the desorption of CO from meteorite nanoparticles (Mautner et al. 2006), such a substrate is likely to be more reproducible in the laboratory. The surface will also be chemically much less complex, allowing a more fundamental approach to the study of adsorption on the grain mimic surface to be adopted.

Evidence suggests that $C_6H_6$ is readily destroyed in diffuse regions by interaction with protons and UV photons (Ruiterkamp et al. 2005). However, in denser regions and circumstellar envelopes where molecules are more shielded from the interstellar radiation field, the $C_6H_6$ lifetime is expected to be much longer. Evidence for this has come from the ISO mission with the observation of $C_6H_6$ in the carbon rich protoplanetary nebula (PPN) CRL 618, accompanied by, $C_4H_2$ and $C_6H_2$ (Cernicharo et al. 2005). These observations have suggested that a significant amount of organic chemistry occurs in C-rich proto-planetary nebulae. Various mechanisms have been put forward for the synthesis of $C_6H_6$ in environments exhibiting different physical and chemical conditions. In the dense interstellar medium the formation of hydrocarbon species, including $C_6H_6$, *via* a gas-phase ion neutral scheme involving atomic hydrogen has been proposed (McEwan et al. 1999). A mechanism initiated by reaction of a range of ions with acetylene ($C_2H_2$) has been shown to account for the significantly higher $C_6H_6$ abundance observed in CRL 618 (Woods et al. 2002). A series of reactions based on soot formation involving $C_2H_2$ has also be proposed as a likely route for the synthesis of $C_6H_6$ and PAHs in circumstellar envelopes (Frenklach & Feidelson 1989 ; Cherchneff, Barker & Tielens 1992). It has been suggested that organometallic catalysis in these environments may lead to an increased $C_6H_6$, and therefore PAH, production rate (Ristorcelli & Klotz 1997). In protostellar disks, a different mechanism has been proposed (Woods & Willacy 2007) whereby $C_6H_6$ is formed *via* $c\text{-}C_6H_7^+$ (where *c*- denotes a cyclic species). Following adsorption of this protonated species on a grain surface, proton transfer and charge neutralisation yields adsorbed $C_6H_6$ and H. The efficiency of this proton transfer mechanism on returning $C_6H_6$ to the gas-phase will depend on how strongly the adsorbed $C_6H_6$ is bound to the grain surface. In this work it was noted that values in the existing literature for the desorption energy of a $C_6H_6$ molecule from a bare grain surface can range from 4700 K to as much as 7500 K (Arnett, Hutchinson & Healy 1988 ; Garrod & Herbst 2006 ; Hasegawa & Herbst 1993 ; Lozovik, Popov & Letokhov 1995). In only one of these studies (Arnett, Hutchinson & Healy 1988) is a value derived experimentally, with gas-solid chromatography being used to obtain the desorption energy from graphite, yielding a value of 39±3 kJ mol$^{-1}$ (4750±360 K). It is clear that such an experiment is not the most appropriate on which to base an understanding of adsorption to interstellar grains.

There have been many experimental studies of the adsorption of $C_6H_6$ on both single crystal and polycrystalline metal surfaces (See, for example, Jakob & Menzel 1989 ; Haq & King 1996 ; Rockey, Yang & Dai 2006). On many substrates, $C_6H_6$ has been found to adsorb plane parallel to the substrate, bonding *via* the $\pi$ electrons. Amorphous silica is widely used as a support for catalysts and has been studied extensively. It is well known that exposure of silica surfaces to $H_2O$ results in reaction to form surface hydroxyl (OH) groups known as silanols. The resulting hydrophilic surface has been studied recently by *ab inito* methods (Tielens et al. 2008). $H_2O$ was shown to interact with three binding sites with adsorption energies of 44, 46 and 50 kJ mol$^{-1}$, characteristic of hydrogen bonding interactions. $C_6H_6$ has been shown to form weak hydrogen bonds to water *via* the aromatic ring (Suzuki et al. 1992), indicating that such bonds may also be important in the adsorption of $C_6H_6$ on a hydroxylated $SiO_2$ surface.

## 2. EXPERIMENTAL METHODS

The experiments described here were conducted in a stainless steel ultrahigh vacuum (UHV) chamber that has been described in detail elsewhere (Oakes 1994). The chamber is pumped by a combination of liquid nitrogen trapped diffusion pumps and a titanium sublimation pump. Following bakeout to remove water adsorbed on internal surfaces, a base pressure of $2 \times 10^{-10}$ torr is routinely obtainable. The substrate was a 10 mm diameter, polished stainless steel disc that was attached to molybdenum support posts on an oxygen free high conductivity (OFHC) copper mount by tantalum wires. Experiments were conducted on both the polished stainless steel surface and a thin film of $SiO_2$ deposited on this surface in order to gain some insight into the effect of the $SiO_2$ on adsorption. The $SiO_2$ films were deposited in a separate chamber by electron beam evaporation of bulk material. The substrate was maintained at room temperature during the evaporation, and ~200 nm of $SiO_2$ were deposited as measured by a quartz crystal microbalance situated close to the substrate. The film thickness of 200 nm was chosen in order to provide a film thick enough that pathways to the stainless steel substrate are limited without introducing interference effects into any future infrared studies. The substrate was cleaned by heating to 500 K for 15 minutes before cooling, prior to conducting experiments.

The substrate was cooled to a base temperature of around 115 K by a liquid nitrogen reservoir in thermal contact with the sample mount. This temperature is significantly higher than that of dust grains in dense clouds (10-20 K), however, preliminary experiments on a system with a substrate base temperature of 10 K (not shown) have demonstrated that $C_6H_6$ does not desorb below 120 K. Similarly, previous experiments (Bolina, Wolff & Brown 2005, Fraser et al. 2001) have also shown that $H_2O$ does not desorb below the base temperature used in the experiments described here. The substrate could be heated resistively by passing current through the support wires. A K-type (Chromel-Alumel) thermocouple, spot-welded to the edge of the substrate, was used for temperature monitoring. A programmable controller (Eurotherm) was used to provide a linear heating ramp ($\beta$) of $0.1 \pm 0.02$ K s$^{-1}$ during temperature programmed desorption (TPD) experiments.

UV-Spectroscopy grade $C_6H_6$ (Fluka) and deionized $H_2O$ were further purified by repeated freeze-pump-thaw cycles on a dedicated stainless steel preparation line pumped by a diffusion pump. To avoid cross-contamination, the $C_6H_6$ and $H_2O$ were purified on independent manifolds, each with a fine leak valve for dosing into the main UHV chamber. Layers of $C_6H_6$ and $H_2O$ were deposited onto the substrate by backfilling the chamber to a pressure as measured by an uncalibrated ion gauge. Exposures are reported in Langmuir (1 L = 1 × 10$^{-6}$ torr s). Though it is not possible to accurately determine coverages with our experimental arrangement, the surface concentration of adsorbed molecules was estimated by using simple collision theory, assuming a sticking probability of unity and an ion gauge sensitivity correction factor of 6 for $C_6H_6$ (Waddill & Kesmodel 1985).

The desorption of $H_2O$ and $C_6H_6$ molecules during TPD experiments was monitored using a quadrupole mass spectrometer (QMS) (modified VG Micromass PC300D) with a cross-beam ion source and channel electron multiplier (CEM) detector that was operated in analogue (current) mode. The QMS was housed in a differentially pumped stainless steel chamber within the main UHV chamber. This arrangement reduces the background signal from residual gases within the main chamber. The QMS chamber incorporated a small tube in line with the ionization region. During TPD experiments the substrate was moved close to the opening of this tube, which provided a line-of-sight to the ionization region. This significantly reduced the detection of molecules desorbing from other surfaces such as the mount and support wires. $C_6H_6$ and $H_2O$ were detected by their parent ions with masses 78 and 18 m$_u$ respectively. The thermocouple voltage was recorded simultaneously by the QMS software following pre-amplification and subsequently converted to temperature using a previously obtained calibration curve.

### 3. RESULTS & DISCUSSION

In order to assess the nature of the amorphous silica substrate used in these experiments, the surface morphology of the film was investigated using atomic force microscopy (AFM) prior to mounting in the UHV chamber. AFM images were obtained with a silicon nitride tip operated in constant force contact mode. At present there is no detailed understanding of the morphology of interstellar grains. A reasonable approximation can be made by considering that of interplanetary dust particles (IDPs). Whilst it is important to recognise that IDPs are likely to have been processed during the formation of the solar system, we consider only the overall structure of the surface, neglecting any chemical modifications which are not represented in our grain mimic. Figure 1 shows a 20 × 20 μm region of the amorphous silica surface along with an SEM image of an IDP[1]. For clarity the AFM image has been exaggerated in the vertical direction. The amorphous silica surface appears to reproduce well the morphology of the IDP on a similar length scale. The likely presence of pores on grain surfaces, which are not present with the polycrystalline Au substrate used in our earlier studies, may have important consequences for the morphology of adsorbates.

---

[1] IDP image obtained from Wikipedia with permission from E. K. Jessberger. Retreived from http://en.wikipedia.org/wiki/Image:Porous_chondriteIDP.jpg on 12/06/2008

Figure 2 shows TPD traces for the desorption of $C_6H_6$ from the bare amorphous silica surface for low, intermediate and large exposures of $C_6H_6$. In all cases the $C_6H_6$ was deposited whilst the substrate was at its base temperature of 115 K and the TPD traces obtained with a heating rate, $\beta$, of 0.1 K s$^{-1}$ to a temperature of 210 K, above which no further $C_6H_6$ desorption was detected. At the smallest exposure a single desorption feature (Peak A) is present at around 180 - 190 K. As the exposure is increased, this peak moves to lower temperature whilst retaining the same high temperature tail. Peak A appears to saturate between 0.5 and 1 L with further exposure leading to the appearance of a much sharper peak (Peak B) centred around 140 K. This peak continues to shift to lower temperature, though to a lesser extent than that observed for Peak A. As will be discussed, we attribute Peak A to a combination of $C_6H_6$ desorption from the surface of the $SiO_2$ and from within pores in the film. Peak B can be attributed to desorption of $C_6H_6$ from a more crowded layer formed on the surface of the $SiO_2$ film when filling of the pores is complete. At around 3 L, Peak B saturates and Peak C begins to grow at around 140 K. This peak is attributed to desorption from the first few multilayers of the solid $C_6H_6$ ice. It is clear that the leading edges of this peak are not coincident as would be exhibited by zero order desorption. This can be interpreted in terms of a fractional order desorption process. This may arise either when islands of $C_6H_6$ form upon the first layer or as a consequence of the inherent roughness of the underlying $SiO_2$ film. It is also interesting to note that as this peak grows, Peak B is reduced indicating that desorption of the $C_6H_6$ monolayer cannot occur until multilayer desorption begins. Beyond around 10 L a single desorption peak (Peak D) dominates the TPD traces with no monolayer peak being resolved. This peak displays coincident leading edges for all exposures up to the maximum investigated (500 L) and can be attributed to zero order desorption from bulk $C_6H_6$ ice.

The desorption from the lowest exposures of $C_6H_6$ reflects the nature of the surface/adsorbate interaction. Typically, desorption from a sub-monolayer coverage on a flat, well ordered surface will yield a narrow, slightly asymmetric peak, characteristic of first order desorption kinetics. Here the desorption shows a broad tail to high temperature which reflects the rather more complicated nature of the $SiO_2$ surface. This tail can be interpreted in two ways. If the surface, rather than having one type of binding site with a characteristic desorption energy, has a distribution of binding sites, a molecule will preferentially adsorb in the sites with higher binding energies. As the coverage is increased, binding sites with progressively lower binding energies will be occupied. Desorption will occur from those sites with the lowest binding energy first, with the desorption profile reflecting the distribution of binding energies. However, whilst a small distribution of binding energies on this surface might be expected, it is unlikely that this would lead to $C_6H_6$ desorption over such a broad range of temperatures. An alternative explanation is that the observed tail is due to the presence of pores in the $SiO_2$. This behaviour, where the peak gradually shifts to lower temperature with increasing coverage has previously been observed in the desorption of CO from a highly porous $H_2O$ film deposited at 10 K (Collings et al. 2003). This was interpreted as being due to molecules bound within pores desorbing and then re-adsorbing elsewhere on the pore surface. It may then take several of these desorption, re-adsorption steps until the $C_6H_6$ desorbs from the $SiO_2$ substrate into the UHV chamber itself, resulting in a much broader desorption profile. In order to investigate this further, experiments were performed where the surface was first exposed to a small amount of $H_2O$. $C_6H_6$ was then deposited to an exposure of 0.5 L. Figure 3 shows TPD traces for the desorption of 0.5 L of $C_6H_6$ from bare $SiO_2$ and $SiO_2$ that had been pre-exposed to small exposures of $H_2O$. If the broad tail of Peak A were due to a wide distribution of binding energies, then at low coverages $H_2O$ would be expected to bind preferentially at those sites with the highest binding energy. The $C_6H_6$ would then be adsorbed in sites with lower binding energies resulting in a narrowed desorption tail, *i.e.* one that is reduced in intensity at higher desorption temperatures. However, in these experiments the tail is reduced at lower desorption temperatures with the development of a two peak profile. This indicates that Peak A is not due to a significant distribution of binding energies. As the amount of $H_2O$ to which the $SiO_2$ surface is pre-exposed is increased, Peak A is reduced to progressively higher desorption temperatures. As Peak A is reduced, the $C_6H_6$ desorption comes to be dominated by a peak similar to Peak B in Figure 2. These observations can be explained by considering the relative mobilities of $C_6H_6$ and $H_2O$. At the substrate temperature used in these experiments $C_6H_6$ will be more mobile than $H_2O$ upon adsorption and will adsorb throughout the pore network when $H_2O$ is not present. On the other hand $H_2O$, upon adsorption, will be relatively immobile as a consequence of strong hydrogen bonding to $SiO_2$ surface hydroxyl groups, and will not penetrate as deeply into the pore network as $C_6H_6$. $C_6H_6$ dosed after small amounts of $H_2O$ will therefore still be able to adsorb deep within the pores, with the desorption profile displaying the high temperature tail. The stronger interaction between $H_2O$ and the $SiO_2$ surface means that pre-deposited $H_2O$ cannot be displaced by $C_6H_6$. The total number of sites available for $C_6H_6$ adsorption within the pore network will therefore be reduced, particularly near the surface, resulting in the observed reduction in the desorption profile from lower temperatures. Exposure to larger amounts of $H_2O$ will block the pathways to the pore network resulting in $C_6H_6$ adsorbing preferentially on the parts of the $SiO_2$ film that remain exposed, rather than on the surface of the pre-adsorbed $H_2O$. This gives rise to the appearance, at higher water pre-exposures, of Peak B which was previously attributed to the desorption of $C_6H_6$ from the surface of the $SiO_2$ film. The non-wetting behaviour of $C_6H_6$ on $H_2O$ has also been observed in experiments where $C_6H_6$ was adsorbed on a thick layer of $H_2O$ sufficient to completely cover the $SiO_2$ surface (data not shown). Multilayer-like desorption was observed for all exposures, indicating the formation of isolated islands of $C_6H_6$ dominated by $C_6H_6$-$C_6H_6$ interactions on the $H_2O$ ice surface. A detailed study of the interaction of $C_6H_6$ with $H_2O$ ice surfaces will be the subject of a future publication. It is likely that during exposure of the $SiO_2$ film to $C_6H_6$, $C_6H_6$ adsorbs both within the pores *and* on the surface of the film, given that the range of binding energies is small. However, increased mobility during heating may cause surface adsorbed $C_6H_6$ to migrate to within the pore network prior to desorption. Further evidence for the impact of surface roughness and the presence pores comes from comparison with initial TPD experiments performed with the polished stainless steel disc prior to the deposition of the $SiO_2$ film (data not shown). No high temperature tail was observed in the TPD trace, and the monolayer peak saturated at around 0.7 L. Comparing this to the observed monolayer saturation at around 3 L on amorphous $SiO_2$ gives an apparent surface area increase of between 4- and 5-fold. This results both from adsorption in pores, and an increased surface roughness.

In order to obtain values for the kinetic parameters for the desorption processes, a combination of direct analysis of the experimental TPD traces and kinetic modelling were used. To describe the desorption kinetics, the rate of desorption can be written in terms of the surface concentration of the adsorbed species

$$r_{des} = -dN/dt = k_{des} N^n \qquad (1)$$

where $N$ is the surface concentration of the adsorbed species in molecules cm$^{-2}$, $n$ is the desorption order and $k_{des}$ is the rate coefficient, which can be expressed as

$$k_{des} = \nu \exp(-E_{des}/k_B T) \qquad (2)$$

where $\nu$ is the pre-exponential factor, $E_{des}$ is the desorption energy in Joules, and $T$ is the surface temperature in Kelvin. The units of $\nu$ depend on the order, being molecules cm$^{-2}$ s$^{-1}$ and s$^{-1}$ for $n = 0$ and $n = 1$ respectively. If the temperature ramp is linear, as was the case in these TPD experiments, the rate equation can be expressed as (Attard & Barnes 1998)

$$dN/dT = (N^n \nu / \beta) \exp(-E_{des}/k_B T) \qquad (3)$$

where $\beta$ is the heating rate in K s$^{-1}$. If the pumping speed is sufficiently high, as expected in our UHV system, the signal from the QMS is directly proportional to $dN/dT$.

It is clear that in order to obtain the kinetic parameters for a particular TPD trace, the surface concentration of adsorbed molecules is required. We are able to obtain approximate surface concentrations by the application of simple collision theory (Atkins & de Paula 2002) in which the collision frequency, $Z$ in molecules m$^{-2}$ s$^{-1}$, for molecules striking a surface can be expressed as

$$Z = P/(2\pi m k_B T)^{1/2} \qquad (4)$$

where $P$ is the partial pressure of the species being deposited, $m$ is the molecular mass in kg and $T$ is the gas temperature (*i.e.* 298 K). Multiplication by the dosing time yields the surface concentration assuming a sticking coefficient of unity. The pressure reading from the ion gauge must be scaled by the appropriate sensitivity factor, 6 in the case of benzene (Waddill & Kesmodel 1985). Small variations in the dosing conditions can impact on the resulting surface concentration. For this reason, the highest dose of 500 L was taken to be exactly 500 L. The TPD trace for 500 L was integrated to obtain the TPD yield. Actual values were then obtained for the other doses by comparison of TPD yield with that obtained for 500 L. Figure 4 shows how the TPD yield varies with increasing dose. It is clear that there are significant variations at the lowest dose, with actual doses being significantly smaller than expected. Comparison with the highest dose, expected to be the most accurate, is therefore required.

Kinetic modelling was performed using the Chemical Kinetics Simulator (CKS) package[2] (Houle & Hinsberg 1995) to stochastically integrate the differential equations describing the desorption process. We have used this approach previously in the analysis of the desorption of pure water ice (Fraser et al. 2001) and of CO from a water ice matrix (Collings et al. 2003). A simple mechanism for the desorption process was constructed:

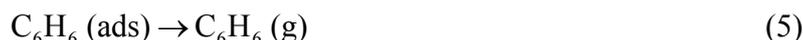

$$C_6H_6 \text{ (ads)} \rightarrow C_6H_6 \text{ (g)} \qquad (5)$$

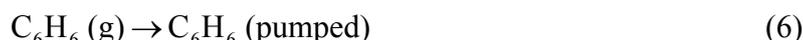

$$C_6H_6 \text{ (g)} \rightarrow C_6H_6 \text{ (pumped)} \qquad (6)$$

where the concentration of $C_6H_6$ (ads) is equivalent to $N$, the surface concentration of adsorbed molecules. Equation (5) describes the desorption step and equation (6) describes the removal of benzene from the UHV chamber by the pumping system, which is proportional to the concentration of $C_6H_6$ (g). With knowledge of the $C_6H_6$ surface concentration, the three variables that remain unknown are the pre-exponential factor for the desorption step, the desorption energy and the pre-exponential factor for the pumping step, hereafter referred to as the pumping speed. Using this model, TPD spectra were simulated for the case of multilayer desorption for raw exposures of 10, 20, 50, 100, 200 and 500 L. Zero order desorption kinetics (*i.e. n*=0) were assumed given the good coincidence of the leading edges. $v$, $E_{des}$ and the pumping speed were systematically varied and resulting simulated traces were compared with the experimental results through use of an appropriate scaling factor. Figure 5 shows a comparison between the experimental and simulated TPD traces for large exposures of $C_6H_6$. The experimental data is best simulated with a $v$ value of $10^{29.7\pm1.0}$ molecules cm$^{-2}$ s$^{-1}$ and an $E_{des}/k_B$ value of 5600±100 K (46.6±0.8 kJ mol$^{-1}$). It is clear that these kinetic parameters describe the experimental TPD traces well over a large exposure range. There is some small deviation at the lowest exposures simulated, which likely results from the desorption order becoming slightly non-zero. This is consistent with the clearly fractional order desorption kinetics observed at exposures of around 10 L, evidenced by non-coincident leading edges. With our approach, it is not possible to obtain reliable results from simulations of fractional order desorption due to the introduction of another unknown variable. The value for $E_{des}$ is in good agreement with reported values for the sublimation energy of condensed $C_6H_6$ which generally lie in the range of 44-47 kJ mol$^{-1}$ (Chickos & Acree 2002) and with previously reported values for the desorption of multilayers of $C_6H_6$ from metal surfaces (see, for example, Jakob & Menzel 1989).

---

[2] Chemical Kinetics Simulator (CKS), Version 1.0, IBM Almaden Research Center, 650 Harry Road, Mailstop ZWX1D1, San Jose, CA, USA. Further information may be obtained from the CKS website at http://www.almaden.ibm.com/st/msim/ckspage.html

It is clear that the desorption of $C_6H_6$ is complicated by the presence of pores, which has a large impact on the desorption profile. A combination of further experimental work and more detailed analysis will be required to fully understand the effect that this porosity has on the desorption process. This is discussed in more detail in the conclusions section of this paper. However, it is possible to obtain a general insight by modelling the desorption using first order desorption kinetics. In this approximation, the steps involved in escaping from the pores are combined into one simple desorption step. Whilst this clearly oversimplifies the desorption process, it is useful in providing a indication of kinetic parameters. Two approaches were used to observe the trends in the desorption energy, $E_{des}$ and the pre-exponential factor, $v$ as a function of $C_6H_6$ exposure. In the first instance, the TPD traces were analysed by assuming a pre-exponential factor of $10^{13}$ s$^{-1}$, as in the frequently invoked Redhead method (Attard & Barnes 1998). Rather than using the Redhead expression directly, the TPD traces were simulated using the kinetic modelling procedure described above to obtain the desorption energy. Our approach was to fit the leading edge of the desorption trace, corresponding to simple first order desorption. The presence of the tail due to desorption from pores makes the determination of the surface concentration minus the tail component difficult. However, by using the pumping speed and scaling factor obtained during the multilayer analysis, the only unknown variables were $E_{des}$ and the surface concentration. These were then systematically varied to obtain the best values. This analysis reveals a wide range of values for $E_{des}/k_B$, decreasing from 6250 K (52 kJ mol$^{-1}$) at the lowest exposure to 4800 K (39.5 kJ mol$^{-1}$) at exposures of around 1 L. The obtained $E_{des}/k_B$ values are shown in Table 1. As discussed previously, we expect the actual range of binding energies to be somewhat smaller than this. Some distribution of binding energies is likely to arise due to increased coordination within the pore network.

Given the complicated nature of the surface, it is highly probable that the assumption that $v$ does not vary with coverage will not hold. For this reason, an independent method was used to obtain values for $E_{des}$. This method has been described previously (Bolina, Wolff & Brown 2005) and is derived by considering that the QMS signal, $I(T)$ is directly proportional to the desorption rate

$$I(T) \propto N^n \nu \exp(-E_{des}/k_B T) \qquad (7)$$

which can be rearranged to yield

$$\ln[I(T)] \propto n\ln[\nu] + n\ln[N] - E_{des}/k_B T \qquad (8)$$

Thus for a given TPD experiment, a plot of ln [$I(T)$] - $n$ ln [$N$] *versus*. 1/T for the leading edge will have a gradient of $-E_{des}/k_B$, from which the desorption energy can be obtained. The value for $E_{des}/k_B$ resulting from this analysis is 5300±200 K (44±2 kJ mol$^{-1}$). In this analysis, this value is independent of coverage in the sub-monolayer regime. The uncertainty on this value is larger than for that obtained for multilayer desorption because of the significantly lower signal to noise ratio in the low coverage regime. This value was then used to obtain corresponding $\nu$ values using appropriate kinetic modelling to reproduce the leading edge of the desorption traces. The values of $\nu$ that were obtained ranged from 4×10$^{10}$ s$^{-1}$ at the lowest exposure to 7×10$^{14}$ s$^{-1}$ at exposures of around 1 L. The obtained $\nu$ values are shown in Table 1. Figure 6 shows a comparison between (a) the experimental TPD traces and (b) those obtained from kinetic modelling. The small values of $\nu$ at low coverages are unphysical and demonstrate the limitations of modelling pore escape as a one step process. At larger coverages $\nu$ increases to values that would be expected for simple first order desorption. The large increase in $\nu$ in going from 0.8 L to 1 L is likely related to the restructuring of the monolayer observed as Peak B in figure 2. Such restructuring, which occurs as a result of repulsive interactions between $C_6H_6$ molecules as the monolayer becomes more crowded, has been observed previously in studies of $C_6H_6$ adsorption on single crystal surfaces. Waddill & Kesmodel (1985) observed the appearance of a TPD peak at lower temperature as the $C_6H_6$ coverage on both Pd(111) and Pd(100) was increased towards monolayer saturation and attributed this to repulsive interactions. Repulsive interactions can have a significant effect on the observed desorption energy of a molecule, in the case of $C_6H_6$ on Pd(111). Monte Carlo simulations (Tysoe et al. 1993) have shown the repulsion energy to be 6.5 kJ mol$^{-1}$ (780 K). Whilst in these experiments the $C_6H_6$ was chemisorbed on the substrate resulting in significantly higher desorption energies than observed in our experiment, where the $C_6H_6$ is more weakly bound, there is no reason to assume that the repulsive interaction would be significantly different. In some cases, for example on polycrystalline Ag (Bahr & Kempter 2007), the increase in intermolecular repulsion has been shown to result in the formation of a second layer where the $C_6H_6$ molecules are adsorbed with their plane tilted with respect to the substrate. This reduces the overlap between the $C_6H_6$ π orbitals and substrate orbitals, resulting in a decrease in desorption energy.

On $SiO_2$, $C_6H_6$ is likely to form weak hydrogen bonds with OH groups on the hydroxylated surface as has been observed in gas-phase $C_6H_6$ - $H_2O$ clusters (Suzuki et al. 1992). However, we have no specific evidence from these experiments to support this. The roughness of the surface means that $C_6H_6$ will not be bound in one particular orientation, rather it will adopt a distribution of orientations that optimises the binding energy. As the coverage is increased, repulsive interactions increase, resulting in some restructuring of the monolayer, resulting in a decreased desorption energy and an increasing pre-exponential factor.

### 4. ASTROPHYSICAL IMPLICATIONS

The experiments outlined here demonstrate the complexity of the interaction between molecules and dust grain surfaces. Surface roughness and the presence of pores have a large impact on the desorption of $C_6H_6$ from such surfaces, particularly at low coverages. The desorption kinetics are affected by interactions between adjacent $C_6H_6$ molecules, the heterogeneous nature of the $SiO_2$ surface, and the presence of pores within the film. These results indicate that the nature of $C_6H_6$ adsorption is likely to be sensitive to the astrophysical environment in which the grain is situated.

We first consider grains in the dense ISM where $C_6H_6$ abundance is expected to be relatively low due to destruction by interaction with UV photons. The distribution of adsorbed $C_6H_6$ will depend on the $C_6H_6$ mobility, and therefore the grain temperature. As grains in the ISM tend to be at the lowest temperatures, $C_6H_6$ mobility will be reduced making it unlikely that $C_6H_6$ molecules will penetrate as far into the pore network as they would at high temperatures when mobility is much higher. The TPD experiments described here probe the high mobility regime as molecules will become mobile during warm up. This same migration due to increased mobility will also occur as grains warm, though the extent to which molecules penetrate the pore network will depend on the presence of any other species that may block access to the pores. The relative mobilities of these species will also be important as observed through the experiments where the surface was first exposed to $H_2O$. Such a situation would occur in the dense ISM where significant amounts of $H_2O$ ice tend to be adsorbed onto grain surfaces. Under these circumstances mobile $C_6H_6$ is unable to penetrate the pore network if the $H_2O$ has previously been sufficiently mobile to penetrate, or there is sufficient $H_2O$ to block access to the pores. The resulting desorption would then be dominated by the kinetics observed at exposures of around 1 L in these experiments, where intermolecular repulsion plays an important role.

This behaviour will also have important consequences for the return of $C_6H_6$ to the gas-phase following formation in protostellar disks. The efficiency of this process will depend on the residence time of $C_6H_6$ on the grain surface which will in turn depend on the grain temperature and the $C_6H_6$-grain interaction. However, whilst we expect the binding energy between $C_6H_6$ and a silicate grain surface to be fairly uniform across the surface, this will be modified by the presence of any pre-adsorbed species. If the grain temperature is sufficiently low to allow formation of a monolayer of $C_6H_6$, or significant $H_2O$ ice is present then intermolecular repulsion will lead to a more efficient return of $C_6H_6$ to the gas-phase than would be expected from a simple consideration of grain temperature and binding energy alone. It is also important to note that carbonaceous grains are likely to dominate in such an environment. The interaction between $C_6H_6$ and, for example, a graphitic surface will be significantly different, though the arguments presented here relating to intermolecular repulsion and the presence of adsorbed $H_2O$ would be expected to hold.

Whilst absorption in pores has a strong effect on the experimental desorption profiles obtained with a heating rate of 0.1 K s$^{-1}$, it is likely that this is not the case at astronomical heating rates. We are currently developing a simple kinetic model to describe the desorption from pores in terms of the mechanism described previously. This model will be based on our previous model for the desorption of CO from porous $H_2O$ films (Collings et al. 2003). Initial results suggest that as the heating rate is lowered the desorption profile tends towards simple first order kinetics. This can be interpreted as the desorption from the surface becoming the rate limiting step, rather than the escape from pores. It may therefore be most appropriate consider the kinetic parameters obtained for the high coverage regime (~1 L) to be most suitable in astrophysical environments.

While the abundance of $C_6H_6$ in astrophysical environments is unlikely to be sufficient for the formation of thick ice layers on grain surfaces, the importance of the multilayer measurements reported here should be highlighted. The kinetic parameters derived are related to the interactions between $C_6H_6$ molecules, rather than between $C_6H_6$ molecules and the underlying substrate. Such interactions become important in the first few layers beyond the monolayer. Furthermore, as discussed, $C_6H_6$ tends to form islands on a $H_2O$ ice surface, with the interactions within these islands more closely resembling those in the multilayer regime.

## 5. CONCLUSIONS

We have reported the results of a series of TPD experiments investigating the adsorption of $C_6H_6$ on a dust grain mimic based on amorphous $SiO_2$. The desorption kinetics of both substrate-bound and multilayer $C_6H_6$ have been studied in detail through a combination of direct analysis and kinetic modelling using stochastic integration of the differential equations describing the desorption process. The desorption of $C_6H_6$ multilayers from adsorbed bulk $C_6H_6$ ice displays simple zero order kinetics with a desorption energy of 5600±100 K and a pre-exponential factor of $10^{29.7\pm1.0}$ molecules cm$^{-2}$ s$^{-1}$, in good agreement with previous studies. However, the desorption of small amounts of $C_6H_6$ from the $SiO_2$ surface demonstrates the effect the underlying surface can have on the desorption process. The presence of pores within the $SiO_2$ and a rough surface, along with the effect of repulsive interactions between $C_6H_6$ molecules leads to complicated desorption kinetics within the first few $C_6H_6$ layers. At low coverages the desorption of $C_6H_6$ molecules from within pores has an important influence on the desorption profile. $C_6H_6$ molecules bound within pores are likely to be re-adsorbed several times before desorbing from the surface. This leads to a high temperature tail being superimposed on the peak related to desorption from the $SiO_2$ surface. Increasing $C_6H_6$ coverage towards monolayer saturation leads to increased repulsion between $C_6H_6$ molecules, resulting in a decreased surface binding energy and an increased pre-exponential factor. It is clear that in order to gain a better understanding of the adsorption on grain surfaces, the coverage dependencies of the kinetic parameters need to be isolated. In particular our kinetic model needs to be modified to specifically include a description of the desorption from pores. Such a model would include both re-adsorption and gas-phase diffusion steps in order to isolate the kinetic parameters for the desorption process. Further insights into the desorption process will be aided by experiments performed on a flat, crystalline $SiO_2$ surface with no pores, which will help to identify the morphological effects of the grain surface on desorption kinetics. This work has demonstrated the sensitivity of the desorption process to both physical and chemical conditions in the local environment.

## ACKNOWLEDGMENTS

JDT and MPC acknowledge the support of the UK Engineering and Physical Sciences Research Council (EPSRC). Financial support from Heriot-Watt University for a number of upgrades to the UHV system is acknowledged. We also acknowledge the support of the Nuffield Foundation for supporting two project students, Stephen McGurk and Sam Zawadzski, to work in our laboratory. We also thank Jekaterina Jeromenok and Claire Pommier for their assistance in the laboratory. We would like to thank the referee for helpful suggestions on improving this manuscript.

# TABLE CAPTIONS

Table 1: Kinetic Parameters obtained for the desorption of $C_6H_6$ from the $SiO_2$ surface. The surface concentrations are those that best reproduce the experimental data based on the pumping speed and scaling factor obtained from the multilayer analysis. Both $E_{des}$ values for a fixed value of $\nu = 10^{13}$ s$^{-1}$ (Redhead model) and $\nu$ values obtained for a fixed $E_{des}/k_B$ value of 5300 K (obtained from leading edge analysis) are shown.

**FIGURE CAPTIONS**

Figure 1: (a) constant force contact mode AFM image of the amorphous $SiO_2$ substrate used in these experiments compared with (b) an SEM image of an interplanetary dust particle. For clarity the AFM image has been exaggerated vertical direction.

Figure 2: TPD traces for $C_6H_6$ desorption from amorphous $SiO_2$. $C_6H_6$ exposures displayed are (a) 0.1, 0.2, 0.5, 0.7, 0.8 and 1 L, (b) 2, 3, 4 and 5 L and (c) 10, 20 and 50 L.

Figure 3: TPD traces resulting from the desorption of 0.5 L $C_6H_6$ from (i) bare amorphous $SiO_2$ and $SiO_2$ that had been exposed to (ii) 0.5 L (iii) 3 L and (iv) 5 L $H_2O$

Figure 4: The TPD yield (integrated peak area), which is proportional to the surface concentration of adsorbed $C_6H_6$, as a function of exposure in L. This reveals a significant overestimate of actual $C_6H_6$ exposure at low coverages as observed by deviation from linearity. Surface concentrations were therefore obtained by comparison of the TPD yield with that for 500 L

Figure 5: (a) TPD traces for $C_6H_6$ multilayer desorption. Exposures are 10, 20, 50, 100, 200 and 500 L. (b) shows the simulated TPD traces that result from a zero order kinetic model with $E_{des} / k_B$ = 5600 K and $\nu = 5 \times 10^{29}$ molecules cm$^{-2}$ s$^{-1}$.

Figure 6: (a) TPD traces for the desorption of $C_6H_6$ from amorphous $SiO_2$. Exposures are 0.1, 0.2, 0.5, 0.7, 0.8 and 1 L. (b) shows the simulated TPD traces first order kinetic modelling. The fixed value of $E_{des} / k_B$ =5300 K was obtained from leading edge analysis of the experimental traces.

**Table 1:**

| $C_6H_6$ Exposure / L | $C_6H_6$ Surface Concentration / molecules cm$^{-2}$ (excluding pore contribution) | $E_{des}\, k_B^{-1}$ / K ($\nu = 1\times 10^{13}$ s$^{-1}$) | $\nu$ / s$^{-1}$ ($E_{des}/k_B = 5300$ K) |
|---|---|---|---|
| 0.1 | $(1.0\pm 0.5)\times 10^{12}$ | 6250 | $4.0\times 10^{10}$ |
| 0.2 | $(2.0\pm 0.5)\times 10^{12}$ | 5890 | $3.0\times 10^{11}$ |
| 0.5 | $(4.0\pm 0.5)\times 10^{12}$ | 5290 | $1.1\times 10^{13}$ |
| 0.7 | $(7.0\pm 0.5)\times 10^{12}$ | 4990 | $1.0\times 10^{14}$ |
| 0.8 | $(9.5\pm 0.5)\times 10^{12}$ | 4930 | $1.5\times 10^{14}$ |
| 1 | $(1.9\pm 0.5)\times 10^{13}$ | 4750 | $7.0\times 10^{14}$ |

**Figure 1:**

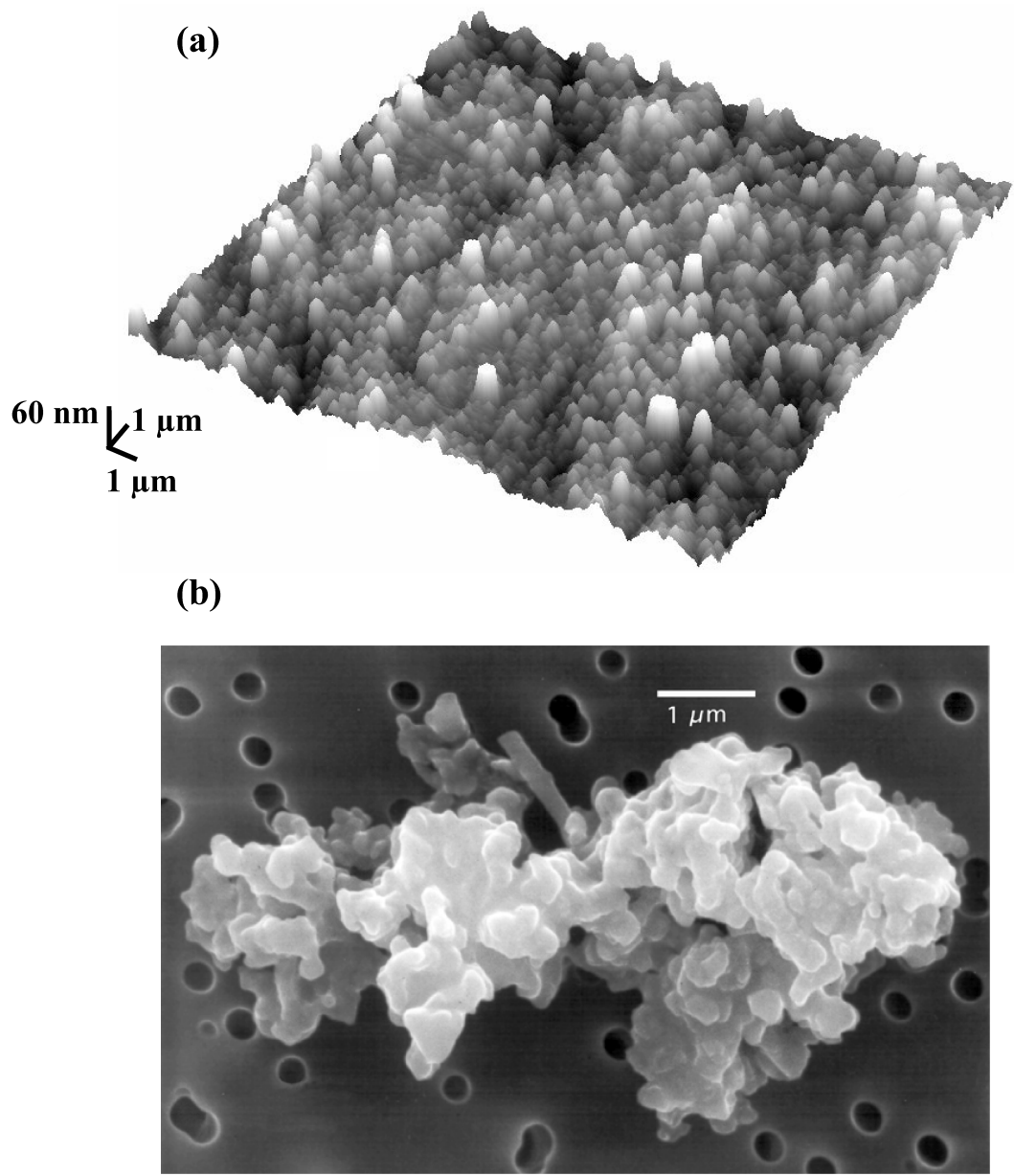

(a)

60 nm, 1 μm, 1 μm

(b)

1 μm

**Figure 2:**

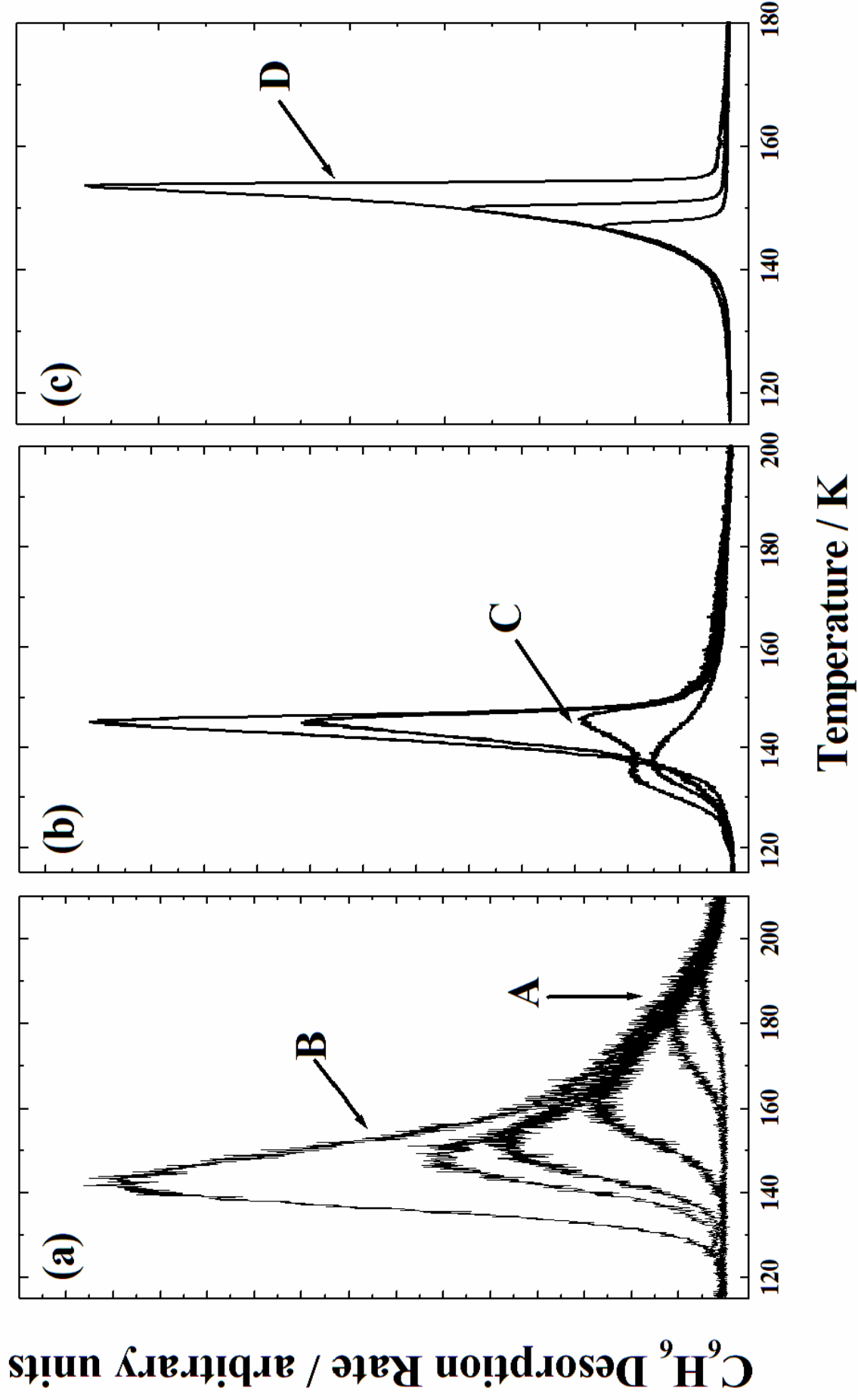

C₆H₆ Desorption Rate / arbitrary units
Temperature / K

**Figure 3:**

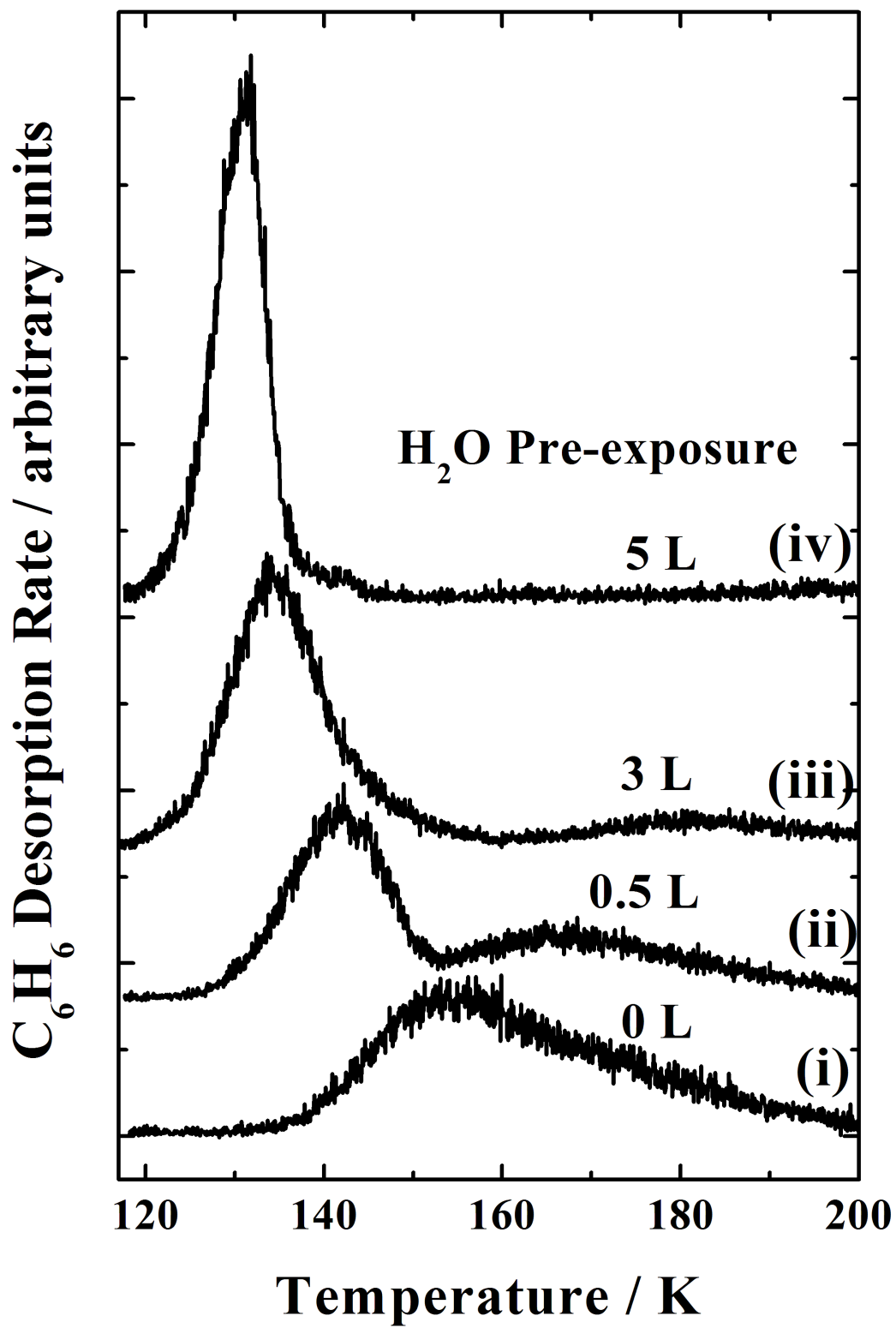

**Figure 4:**

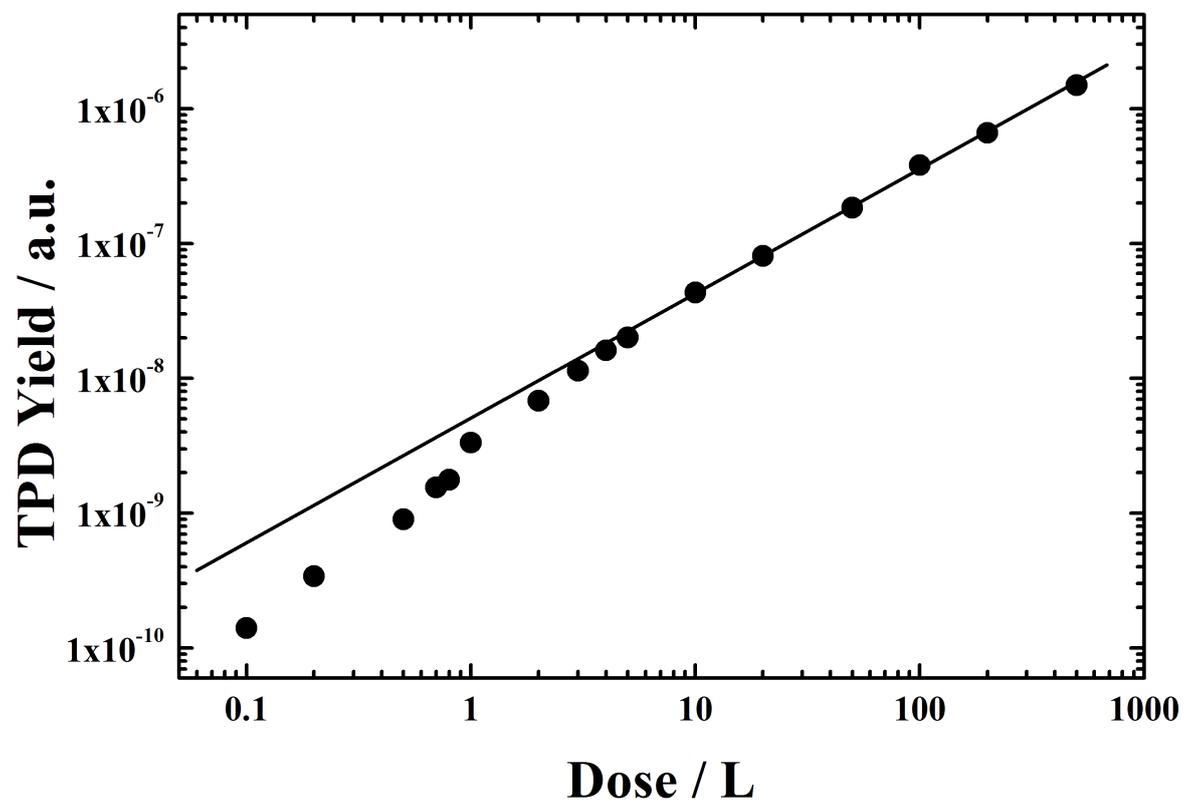

Figure 5:

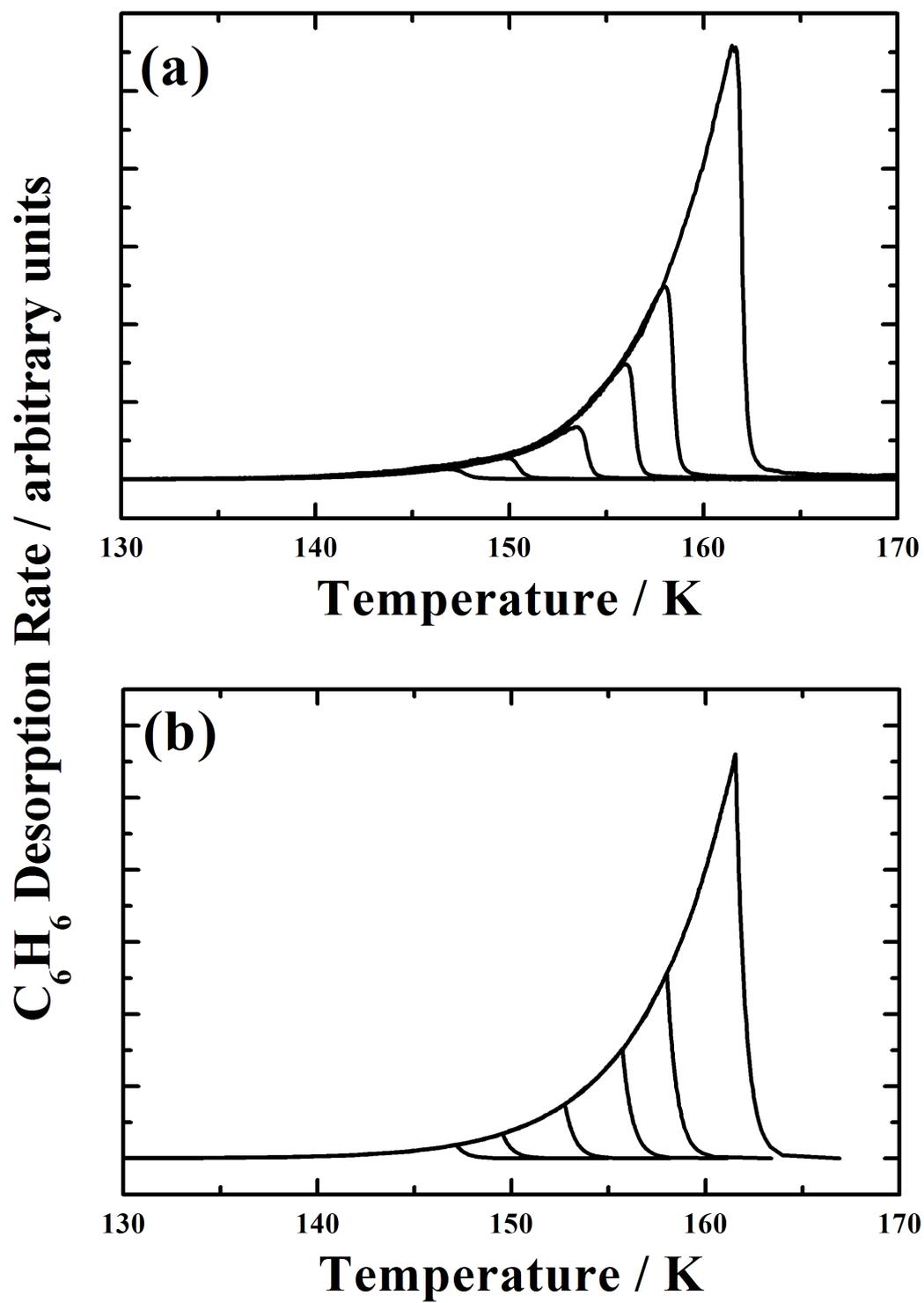

**Figure 6:**

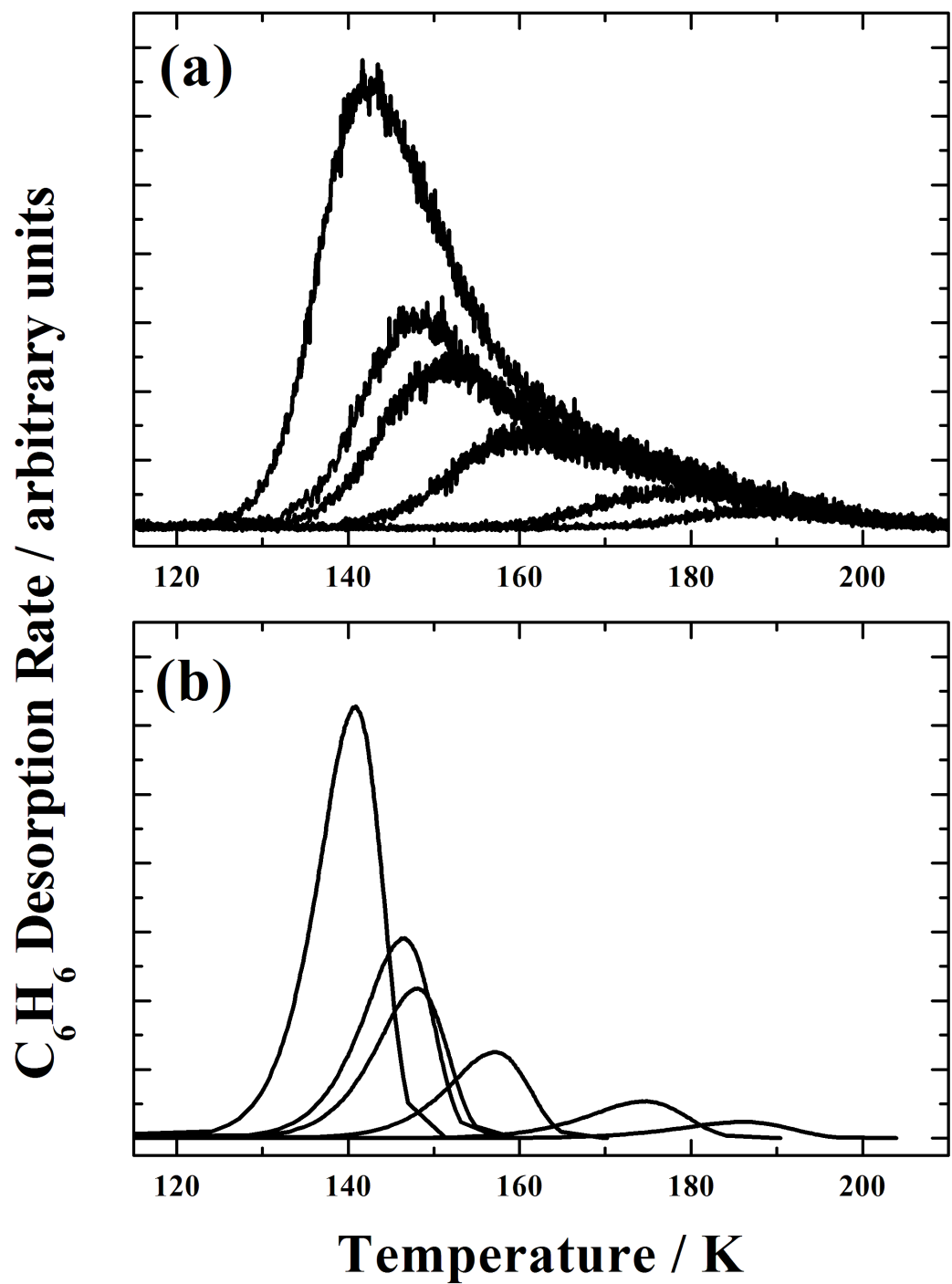